\documentclass[pdflatex]{nature}%
\usepackage{soul,xspace}
\usepackage{bm}
\usepackage{graphicx}%
\usepackage{multirow}%
\usepackage{amsmath,amssymb,amsfonts}%
\usepackage{amsthm}%
\usepackage{mathrsfs}%
\usepackage[title]{appendix}%
\usepackage{xcolor}%
\usepackage{textcomp}%
\usepackage{manyfoot}%
\usepackage{booktabs}%
\usepackage{algorithm}%
\usepackage{algorithmicx}%
\usepackage{algpseudocode}%
\usepackage{listings}%
\usepackage[font=footnotesize]{caption}
\usepackage[unicode, pdftex, colorlinks=true]{hyperref}
\usepackage[%
separate-uncertainty=true,%
print-zero-exponent=false,
print-unity-mantissa=false
]{siunitx}
\def\nat{Nature\ }
\def\aap{Astron.\ Astrophys.\ }
\def\apj{Astrophys.\ J.\ }

\def\apjs{Astrophys.\ J.\ Supp.\ }

\def\mnras{Mon.\ Not.\ Roy.\ Astron.\ Soc.\ }
\def\physrep{Phys.\ Rept.\ }
\def\prd{Phys.\ Rev.\ D\ }
\def\prl{Phys.\ Rev.\ Lett.\ }

\def\jcap{J.\ Cosmol.\ Astropart.\ Phys.\ }

\def\lat{\textit{Fermi/LAT}\xspace}

\def\ga{g_{a\gamma}}

\newcommand{\beq}{\begin{equation}}
\newcommand{\eeq}{\end{equation}}
\newcommand{\bea}{\begin{eqnarray}}
\newcommand{\eea}{\end{eqnarray}}

\title{Active galactic nuclei through the prism of galaxy clusters: bounds on axion-like particles}

\begin{document}

 \author{Denys Malyshev$^{1}$, 
 Lidiia Zadorozhna$^{2,3}$, 
 Yuriy Bidasyuk$^{4}$,
 Andrea Santangelo$^{1}$,
 and Oleg Ruchayskiy$^{2}$}

\maketitle

 $^1$Institut f{\"u}r Astronomie und Astrophysik T{\"u}bingen, Universit{\"a}t T{\"u}bingen,  Sand 1, D-72076, T{\"u}bingen, Germany

 $^2$Niels Bohr Institute, Jagtvej 155A, 2200, K{\o}benhavn, Denmark

 $^3$Taras Shevchenko National University of Kyiv, Hlushkova ave. 4, 03127, Kyiv, Ukraine

 $^4$Bogolyubov Institute for Theoretical Physics
 of the National Academy of Sciences of Ukraine, Metrolohichna str. 14-b, 03143, Kyiv, Ukraine\\

\begin{abstract} 
Hypothetical axion-like particles (ALPs) are of interest because of their potential to act as dark matter or to reveal information about yet undiscovered fundamental constituents of matter.
Such particles can be created when photons traverse regions of magnetic fields.
The conversion probability depends on both the magnetic field parameters and photon energy, leading to multiple spectral absorption features as light passes through magnetized regions.

\medskip
Traditionally, astrophysical searches have focused on detecting such features in individual objects. 
However, the limited understanding of properties of cosmic magnetic fields have hindered the progress.
Here we introduce a new approach by analyzing \emph{stacked} (rather than individual) spectra of active galactic nuclei (AGNs) positioned behind galaxy clusters -- gigantic magnetic field reservoirs. 
Stacking efficiently averages over the uncertainties in magnetic fields, revealing a unique step-like spectral signature of photon-to-ALP conversion.

\medskip

With this approach we advance into previously inaccessible regions of the ALP parameter space for 
nano-electronvolt masses. 
Adopting this method will significantly improve existing bounds across a wide range of masses by using different telescopes and increasing the size of the stacked datasets.
The Cherenkov Telescope Array Observatory, in particular, will extensively probe the parameter space where ALPs could serve as dark matter.

\end{abstract}




\maketitle

\bigskip

Cosmic environments provide unique physical conditions (densities, temperatures, spatial and temporal scales) unattainable in terrestrial experiments. Consequently, astrophysical sites can serve as alternative laboratories for particle physics. In particular, they offer unique opportunities to study hypothetical particles that interact minimally with known forms of matter\cite{1996slfp.book.....R}. 
Extreme astrophysical conditions may facilitate the production of these particles. Thus their presence can be deduced from astronomical observations.

Many extensions of the Standard Model of particle physics predict the existence of new, currently undetected, particles.
These extensions are motivated by observational challenges such as the existence of dark matter, the absence of primordial antimatter in the Universe, or by quests to understand the underlying structure of physical theories, see Refs.\cite{2005PhR...405..279B, 2000PhR...325....1R, 2022SciA....8J3618C} for a review. 

\emph{Axions} are prime examples of such particles.
Originally postulated in an attempt to solve the so-called strong-CP problem -- a mysterious cancellation between two seemingly unrelated contributions, leading to the absence of charge-plus-parity violation in strong interactions\cite{Peccei:1977ur,Peccei:1977hh} -- they have quickly been recognized as promising dark matter candidates\cite{Wilczek:1977pj,Weinberg:1977ma}.
The subsequent development of the idea led to a broader concept of \emph{axion-like particles} (ALPs) -- light pseudo-scalar bosons with mass $m_a$ whose interaction with electromagnetic fields is governed by the term~\eqref{eq:lagrangian_int_1}
\begin{equation}
  \label{eq:lagrangian_int_1}
  \mathcal{L}_{\rm int} = \ga {\bm E}\cdot{\bm B}\,a\,. 
\end{equation}
Here, $a$ is the ALP field, and $\bm{E}, \bm{B}$ are the electric and magnetic fields, respectively (see \textbf{Methods} for details).
The unknown ALP-photon coupling constant, $\ga$, characterizes the strength of the interaction and is measured in $\si{GeV}^{-1}$.

\begin{figure}[!t]
    \centering
    \includegraphics[width=\linewidth]{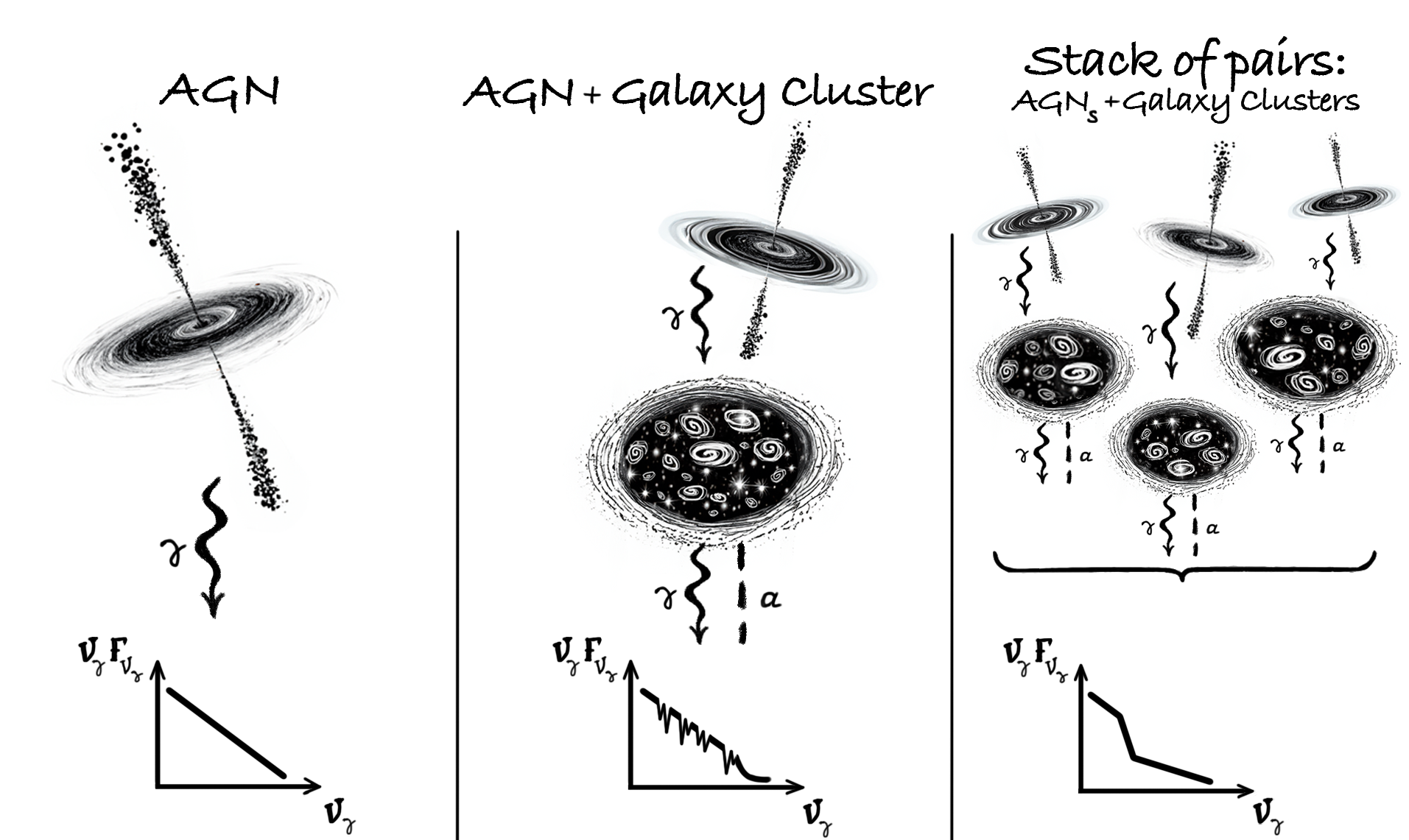}
    \caption{\textbf{Illustration of our method.} Active galactic nuclei have featureless spectra of $\gamma$-ray emission (\textit{left panel}). If axion-like particles exist in nature, some of the AGN photons will be converted to ALPs while passing through galaxy clusters that are large reservoirs of magnetic fields. Such a photon-to-ALP conversion creates a set of absorption features in the AGN spectra. For each particular AGN these features cannot be predicted due to the lack of detailed knowledge about the magnetic field in individual clusters (\textit{central panel}). By stacking a large number of observational pairs (AGNs plus clusters) the absorption feature becomes regularly shaped (\textit{right panel}).}
    \label{fig:cartoon}
\end{figure}

The interaction term given by Equation~\eqref{eq:lagrangian_int_1} causes energy-dependent photon-to-ALP conversion when photons traverse regions with strong and/or large-scale magnetic fields\cite{1996slfp.book.....R}. Specifically, as photons emitted by an astrophysical object pass through such regions, photon-to-ALP conversion results in spectral irregularities (e.g., multiple absorption features) in the observed emission. Detection of these features in otherwise smooth spectra of astrophysical objects would be a smoking gun of the ALPs existence. This approach offers thus a compelling method for identifying ALPs and highlights astrophysical sites like neutron stars\cite{2016PhRvD..93f5044S,Fiorillo:2021gsw}, white dwarfs\cite{1988PhRvD..37.1237R, 2019PhRvL.123f1104D}, active galactic nuclei\cite{2015PhLB..744..375T,Davies:2022wvj} supernovae\cite{1996PhLB..383..439B, 2002PhRvL..88p1302C, 2002PhLB..543...23G,Crnogorcevic:2021wyj} and clusters of galaxies\cite{2016PhRvD..93l3526C, PhysRevLett.116.161101, Marsh:2017yvc, 2020ApJ...890...59R} as prime laboratories for ALP searches.

\begin{figure}[!t]
    \centering
    \includegraphics[width=0.7\textwidth]
    {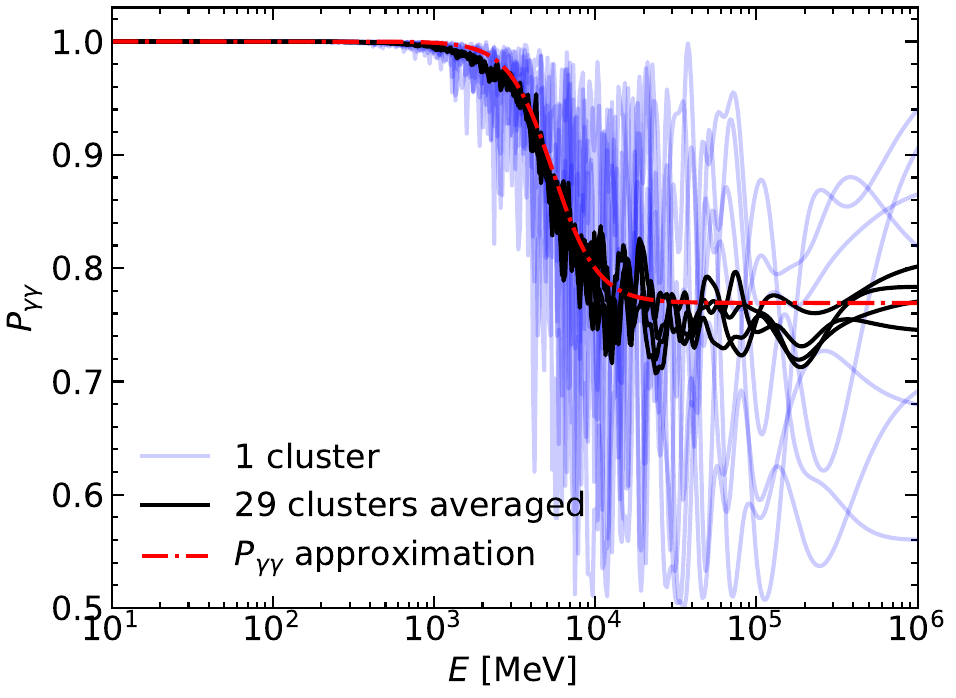}
    \caption{\textbf{Photon survival probability and its averages.} Photon survival probability when passing through the galaxy cluster for different realizations of the cluster magnetic field ({\bf blue lines}). All realizations have the same  radial profile of the magnetic field but vary randomly in their orientation in the photon polarization plane and in the sizes of the domains where the field remains approximately constant. {\bf Black lines} demonstrate the effect of averaging over 29 randomly selected realizations. \textbf{Red dot-dashed line} shows the analytical approximation to these lines, given by the expression~\protect\eqref{eq:p_ga}.
    The ALP parameters for all curves are $(m_a, \ga) =(\SI{3}{neV}, \SI{2e-12}{GeV^{-1}})$. Blue curves were obtained by numerically solving ALP propagation equations via the \texttt{ALPro} code\cite{Matthews2022}.
}
    \label{fig:conversion}
\end{figure}

The search for ALP-induced absorption features in individual bright astrophysical sources typically employs a statistical approach to detect spectral irregularities. The specific characteristics of these spectral features are determined by the exact distribution of the intervening magnetic field. 
Even minor variations in the magnetic field's orientation, spatial distribution, or strength affect the photon-to-ALP conversion probability in a quasi-random way, as illustrated in Figures~\ref{fig:cartoon} and \ref{fig:conversion}. 
The existing methods therefore require marginalizing over a large number of potential realizations of random magnetic fields within a selected object, aiming to address the substantial uncertainties and poor knowledge of the magnetic field\cite{2018SSRv..214..122D, 2020ApJ...890...59R, Kachelriess:2021rzc, Davies:2022wvj, Cecil:2023iU,
2007PhRvD..76l3011H, Wouters_2013, Marsh:2017yvc, Conlon:2017qcw, 2018arXiv180504388M, 2022MNRAS.510.1264S, Matthews2022, 2023PhRvD.107h3027D
}. 
Therefore, the improvement of astrophysical bounds on ALP couplings has been hindered by the challenge of accounting for all the unknowns. 

\begin{figure}[!t]
    \centering
    \includegraphics[width=\textwidth]{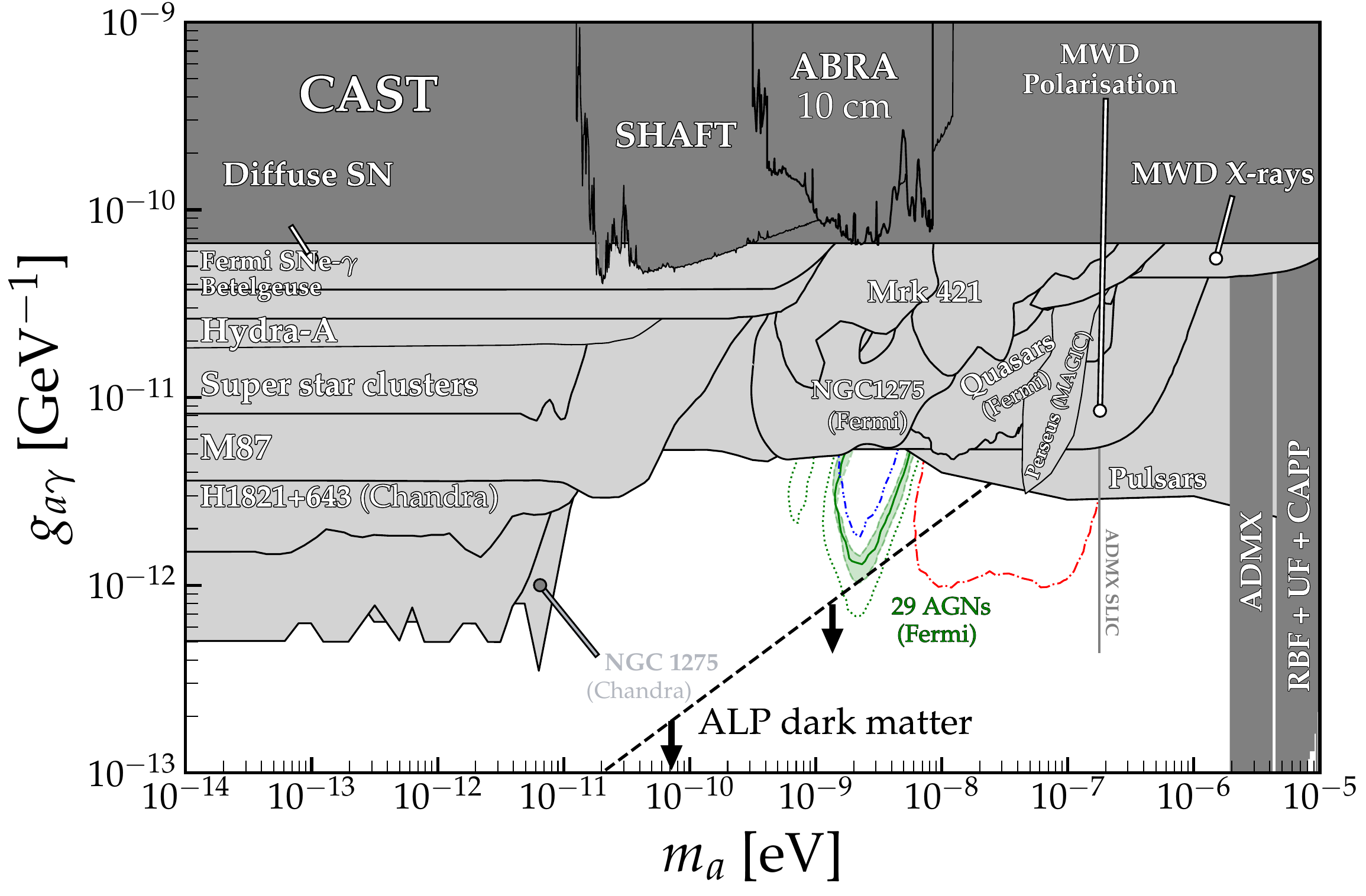}
    \caption{\textbf{ALP-photon coupling constraints from AGN-cluster pairs.}
The green solid curve represents the 95\% upper limits on the coupling $\ga$ derived from the stacked analysis of AGN spectra behind galaxy clusters, based on our estimate of the average magnetic field across the sample.
The green-shaded region surrounding this curve indicates the uncertainty in the field estimate.
The blue dash-dot-dotted line illustrates a pessimistic scenario where the magnetic field weakens with decreasing cluster mass.
The green dotted line shows constraints based solely on statistical errors.
Between the two disjoint regions, the quality of fit improves by up to $2\sigma$ when including an ALP.
The red dot-dashed line highlights potential improvements achievable with the CTAO using the same AGN-cluster pairs. Gray regions display existing exclusion bounds from Ref.\cite{ciaran3932430}.}
    \label{fig:broad_constraints}
\end{figure}

\subsection{Stacked analysis.}
In this work, we explore a radically different approach, see Figure~\ref{fig:cartoon}. 
Rather than focusing on individual bright sources of photons and marginalizing over the details of the magnetic field distribution within these objects,
 we exploit the power of ensemble averaging over a large number of sources.
Figure~\ref{fig:conversion} illustrates this approach.
The blue curves show the survival probability of a photon as it traversed a region of the magnetic field (e.g., a galaxy cluster), with each curve representing a different realization of the field.
This probability (and the resulting absorption features in the observed spectrum) has a complex shape, with little similarity between different realizations.
However, the averages over a large number of such curves (shown as black lines) are much smoother and show significant similarities.
Moreover, the black curves exhibit a distinct, realization-independent spectral signature -- a step-like suppression at high energies -- which serves as a clear indicator for ALP searches. This makes studying photon-to-ALP conversion in a \textit{stacked set} of clusters substantially simpler than conducting searches in \textit{individual} objects.

To implement this idea, we stack the spectra of $\gamma$-ray bright active galactic nuclei (AGNs) located behind galaxy clusters. AGNs have smooth spectra in the GeV energy range, while galaxy clusters serve as vast reservoirs of magnetic fields.
Consequently, photon-to-ALP conversion may occur, leading to a suppression of the high-energy portions of AGN spectra.

\noindent\textbf{Main steps of the data analysis.}
Using the most recent \lat catalog\cite{2022ApJS..260...53A, 4fgldr4} and the galaxy cluster catalog\cite{2018MNRAS.475..343W} we have identified 29 AGNs that are located behind known galaxy clusters (selection procedure is detailed in the \textbf{Methods} section).
We fit each AGN spectrum with a log-parabola (smooth 3 parametric function, known to describe well the AGN $\gamma$-spectra\cite{2018A&A...614A...6S}). We also account for the interaction of $\gamma$-ray photons with extra-galactic background light (EBL), which produces electron-positron pairs.
Such an interaction leads to the suppression of the high-energy part of the AGN spectrum, partially mimicking the effect of photon-to-ALP conversion. 
The suppression is a function of the AGN redshift only and does not increase the number of fitting parameters. 
This fit is our baseline model.

Subsequently, we repeat the spectral analysis by multiplying each of the above functions by a \textit{common smooth function} (red line in Figure~\ref{fig:conversion}):
\begin{equation}
P_{\gamma\gamma} \equiv 1 - \bigl\langle P_{\gamma a} (E )\bigr\rangle = 1- \frac{p_0}{1 + (E_c/E)^k}.
\label{eq:p_ga}
\end{equation}
By demanding that the resulting $\chi^2$ of the fit worsens by $\Delta \chi^2 \geq 6.2$, we  draw the 95\% confidence level (CL) contours in the plane ($m_a, \ga$), excluding ALPs with the corresponding parameters. 

\subsection{Results.}
The green solid line in Figure~\ref{fig:broad_constraints} defines the excluded region of ALP parameters and represents our main result.
We have strengthened the existing constraints\cite{ciaran3932430} by up to a factor of $4$ for ALP masses in the range of $\SIrange{1}{10}{neV}$.
In deriving this bound, we assumed that all clusters in our sample possess magnetic fields similar to that of the Coma cluster. The green-shaded area around the line reflects how the uncertainty in both the magnitude and spatial distribution of the Coma cluster's magnetic field propagates to the ALP bounds.
The blue dash-dot-dotted line represents an alternative scenario where the magnetic field strength correlates with cluster mass, as suggested by some numerical simulations\cite{2024A&A...686A.157N}.
However, existing observations do not support such scaling\cite{2001A&A...378..777D, 2017A&A...603A.122G, 2012A&ARv..20...54F}, see \textbf{Methods} for details.
Therefore, we consider this scenario overly pessimistic.

These results take into account systematic uncertainties arising from the imprecise knowledge of the \lat effective area. 
Following the collaboration guidelines\cite{fermi_lat_website}, we assume energy-dependent systematics ranging from $3\%$ to $10\%$.
For comparison, we also present a purely statistical bound (the dotted line), derived without including such systematic uncertainty. 
In this case, the constraints on $\ga$ improve the existing bounds by up to a factor of $7.5$.
The improved bounds extend into the region of parameter space where ALP particles could potentially serve as dark matter candidates\cite{2009EPJC...59..557S}.

Intriguingly, a second disjoint exclusion region appears in this case.
Between the two regions, the ALP actually \emph{improves} the fit quality, indicating a marginal detection at approximately $2\sigma$ level.
The shape of this region follows a $\ga \propto m_a^2$ relation, which maintains a constant characteristic energy $E_c$ in Eq.~\eqref{eq:p_ga}.
Upon detailed inspection (see Figure~\ref{fig:chi2_agns} and Table~\ref{tab:detection_table} in \textbf{Methods}), we observe that most AGN spectra show a consistent improvement in fit quality when the characteristic energy $E_c$ is close to $E \sim \SI{600}{MeV}$, as one would expect from true signal detection, rather than an artifact of a single AGN spectrum.
It is worth noting that the ALP parameters that show the maximal $\Delta\chi^2$ improvement are located in the region of parameter space where the existence of such particles has previously been suggested based on TeV transparency arguments\cite{tev_transparency} and stellar evolution studies\cite{cooling_hint}. 

\subsection{Future improvements.}
In this work, we have put forward a new method of searching for axion-like particles by using stacked spectra of AGNs located behind galaxy clusters. Collectively fitting a large number of spectra allows us to search for a regular step-like suppression feature rather than irregular photon transparency in individual clusters. The uncertainties related to the unknown characteristics and distribution of magnetic fields in individual galaxy clusters are reduced to essentially one number -- the average magnetic field across the sample.

Our method not only proposes the most competitive constraints but also shows a great potential for improvement.
The observations of a similar number of sources at other wavelengths with current or future missions, e.g.,\ at keV energies with Swift, at MeV energies with AMEGO and GECCO, and at TeV energies with Cherenkov Telescope Array Observatory (CTAO)\cite{2017arXiv170905434C, 2022icrc.confE.885G}, would result in a substantial extension of the derived limits to lower and higher ALP masses. 
This is illustrated by the red dashed line in Figure~\ref{fig:broad_constraints}, which shows the potential reach of similar searches performed with the CTAO. Different energy ranges of the CTAO open access to larger ALP masses and promise an order of magnitude improvement in $\ga$.
In particular, such searches will explore a part of the ALP dark matter parameter space.

Our current analysis is based on only 29 AGN-cluster pairs. 
Further progress will be achieved as the number of galaxy clusters detected via the Sunyaev-Zeldovich effect increases with present-day X-ray all-sky surveys such as eROSITA\cite{2024A&A...682A..34M}.

Future studies should also consider several aspects beyond this proof-of-concept work. These include detailed investigations of the magnetic field distribution in galaxy clusters and the development of a physically motivated model of the background AGN emission. Additionally, we believe that studying average magnetic fields across multiple galaxy clusters may be more manageable than examining the magnetic field in detail within a single cluster.

\clearpage 
\section*{Methods}
\subsection{ALP propagation equations.}
In this Section, we summarize the key theoretical ingredients relevant to the studies of ALPs, and photon-to-ALP conversion in external magnetic fields. An extensive review of the subject can be found in, e.g., Ref.\cite{2021RvMP...93a5004S}.

Interactions of axion-like particles with electromagnetic field are governed by the following Lagrangian:
\begin{equation}
    \label{eq:lagrangian}
    \begin{aligned}
  {\cal L}=-\frac{1}{4}
  F_{\mu\nu}{F}^{\mu\nu}+\frac{1}{2}\left(\partial_{\mu}a\partial^{\mu}a-m_{a}^2a^2\right)+\frac{1}{4}\ga F_{\mu\nu}\tilde{F}^{\mu\nu}\,a.
\end{aligned}
\end{equation}
Here, $a$ is the ALP field, $m_a$ is its mass, $F_{\mu\nu}$ is the electromagnetic field strength tensor, and $\tilde{F}_{\mu\nu}\equiv\frac{1}{2}\varepsilon_{\mu\nu\rho\sigma} F^{\rho\sigma}$ is the electromagnetic dual tensor. The photon-ALP coupling constant, $\ga$, characterizes the interaction strength. The coupling between ALPs and electromagnetic fields, described by the third term in Eq.~(\ref{eq:lagrangian}), can be expressed as in Eq.~(\ref{eq:lagrangian_int_1}). We use natural Lorentz-Heaviside units with $\hbar=c=k_{B}=1$ and the fine-structure constant $\alpha=e^2/4\pi\approx1/137$, where $e$ is the electron charge.

Lagrangian~\eqref{eq:lagrangian} suggests that a photon can convert into an ALP when passing through a magnetic field\cite{1988PhRvD..37.1237R}.
To derive the relevant formulas, consider a photon with energy $E$, propagating through the magnetic field in the $z$ direction. 
The component of the magnetic field perpendicular to the propagation direction is denoted by ${\bm B}_{\perp}={\bm B}-B_z{\bf e}_z$.
Joint evolution of the perpendicular photon components $(A_x, A_y)$ and ALP $a$ are described by the equation\cite{1988PhRvD..37.1237R,
2007PhRvD..76l3011H, 2009JCAP...12..004M}
\begin{equation}
\label{eq:propagation}
\Biggl[ E - i\frac{\partial}{\partial z} - \mathcal M(m_a, \ga, \bm{B}_\perp(z) \,) \Biggr]\left(\begin{array}{l}A_x \\ A_y \\ \,a \end{array}\right) = 0, 
\end{equation}
where the mixing matrix $\mathcal{M}(m_a, \ga, \bm{B}_\perp(z))$ depends on the magnetic field's strength and orientation, as well as on the ALP mass $m_a$ and the coupling constant $\ga$.
For MeV-GeV photon energies the matrix $\mathcal M$ is given by:
\begin{equation}
\label{eq:mixing_matrix}
    \mathcal M = \left(\begin{array}{ccc}
    0 & 0 & \Delta_{a\gamma}\cos\phi \\
    0 & 0 & \Delta_{a\gamma}\sin\phi \\
\Delta_{a\gamma}\cos\phi & \Delta_{a\gamma}\sin\phi & \Delta_a
\end{array}\right),
\end{equation}
where $\cos\phi={\bm B}_{\perp}\cdot{\bf e}_x/B_{\perp}
= \sqrt{1-\sin^2\phi}$ and
\begin{align}
\label{eq:Delta_a}
& \Delta_a = -\frac{m_a^2}{2E}
\simeq 
       -7.04 \times 10^{-4} \left(\frac{m_a}{3\times 10^{-9}\, 
        {\rm eV}}\right)^2 \left(\frac{E}{{1\,\rm GeV}} \right)^{-1} {\rm pc}^{-1}, \\ \nonumber
& \Delta_{a\gamma} = \frac{1}{2} \ga B_{\perp}
\simeq 3.05 \times10^{-6} \left(\frac{\ga}{2\times 10^{-12}\,\rm {GeV}^{-1}} \right)
\left(\frac{B_{\perp}}{1\,\mu\rm G}\right) {\rm pc}^{-1}.
\end{align}
For lower energies, the plasma frequency of the galaxy clusters and the possibility of resonant photon-to-ALP conversion should be taken into account; see, e.g., Refs.\cite{2008LNP...741..115M, Marsh:2021ajy}. 
It is instructive to solve Eq.~(\ref{eq:propagation}) for the scenario where a photon propagates through a region with a constant magnetic field. The probability of photon-to-ALP conversion is energy-dependent, and after traveling a distance $l$, it is given by\cite{1988PhRvD..37.1237R}:
\begin{equation}
\label{eq:p_0}
  P_{\gamma a}\simeq \frac{(\Delta_{a\gamma} l)^2}{(\Delta_{\rm osc} l/2)^2}\sin^2
\left( \frac{\Delta_{\rm osc} \, l}{2} \right)\,.
\end{equation}
Here, $\Delta_{\rm osc}$ -- the oscillation wavenumber -- is given by
\begin{equation}
\label{a17}
{\Delta}_{\rm osc}\simeq 
\left[\Delta_a^2 + 
4 \Delta_{a \gamma}^2 \right]^{1/2}
= 2 \Delta_{a \gamma} \sqrt{1 + \left(\frac{E_c}{E} \right)^2}
 \,\ ,
\end{equation}
and $E_c$ represents the characteristic energy of oscillations in $P_{\gamma a}$:
\begin{eqnarray}
\label{eq:E_c}
{E}_c &\equiv& E
\frac{\Delta_a}{2 \Delta_{a \gamma}}
\simeq  {\SI{115}{GeV}}
\left(\frac{m_a}{\SI{3}{neV}}\right)^2
\left( \frac{B_{\perp}}{1\,\mu\rm G} \right)^{-1}
\left( \frac{\ga}{2 \times 10^{-12}\,\rm GeV^{-1}} \right)^{-1}.
\end{eqnarray}

\subsection{Conversion probability averaged across many domains and objects.}
The above formulas concern an idealized setup of photon-to-ALP conversion in the constant magnetic field.  
More relevant for our discussion is the case when the field changes along the photon's trajectory.
To wit, consider $N$ domains of size $l$, with the amplitude of the magnetic field being the same in each of them. 
The conversion probability in a single domain, $P_{\gamma a}$ is given by Eq.~(\ref{eq:p_0}). 
It can be shown\cite{2002PhLB..543...23G,2008LNP...741..115M, 2008PhLB..659..847D,Kachelriess:2021rzc} that by crossing $N\gg 1$ domains and averaging over the orientations of the magnetic field across many similar objects, the conversion probability simplifies to
\begin{equation}
\label{eq:analytic}
\langle P_{\gamma a}\rangle \simeq \frac{1}{3}\left(1-e^{-\frac{3}{2}NP_{\gamma a}}\right),
\end{equation}
Thus defined, $\langle P_{\gamma a}\rangle$ is a \textit{step-like function} of the photon energy $E$. 
While at low ($E\ll E_c$) energies $\langle P_{\gamma a}\rangle\approx 0$, it saturates to a constant at $E\gg E_c$ with a saturation level proportional to $\ga$ for $\langle P_{\gamma a}\rangle\ll 1/3$ and asymptotically reaching 1/3 with the increase of $\ga$. 

The $\langle \dots \rangle$ symbol stresses that this result appears only after \textit{averaging over many objects}. Without such averaging, we would not obtain a step-like suppression (corresponding to black curves in Figure~\ref{fig:conversion}) but rather would stay with a random blue curve. 

\subsection{Photon-to-ALP conversion in the inhomogeneous magnetic field of galaxy clusters.}
\label{sec:photon-alp-conversion}
Finally, we describe how we obtained the expression $\langle P_{\gamma a}\rangle$ in the case when not only the orientation but also the amplitude of the magnetic field changes along the line of sight. 
For the case of a realistic spatial profile of the cluster magnetic field, we solve Eq.~\eqref{eq:propagation} numerically using the \texttt{ALPro} code\cite{Matthews2022}.
This code accepts as an input the list of magnetic field magnitudes and orientations in a set of domains along the light's trajectory. Within each of these domains, the strength and the orientation of the magnetic field are constant. 
We assume that the strength of the magnetic field in each cluster is proportional to the density of plasma electrons and depends only on the distance to the center of the cluster $r$:
\begin{equation}
\label{eq:magn_f}
  B(r) = B_{0} [n_e(r)/n_0]^{\eta},
\end{equation}
where $B(r)$ is the amplitude of the magnetic field, $B(r) = |\bm{B}(r)|$.
The parameters in Eq.~\eqref{eq:magn_f} are adopted from those of the Coma cluster -- the only galaxy cluster where the magnetic field strength profile has been determined fairly well\cite{2018MNRAS.481.5046R, Bonafede:2010xg}.
The density of electrons is described by the $\beta$-model, 
$n_e(r)/n_0 = [1 + (r/r_c)^2]^{-3\beta/2}$ with 
$\beta=0.75$; and $r_c = 291$~kpc\cite{Bonafede:2010xg,2003MNRAS.343..401L}. 
The values of $(B_0,\eta) = (\SI{5.2}{\mu G},0.67)$ are adopted from Ref.\cite{Bonafede:2010xg}.

The domain sizes are distributed randomly between 2~kpc and 34~kpc, based on Refs.\cite{Bonafede:2010xg, 2015JCAP...01..011K}. 
The cluster radius is set to 1.5~Mpc, as the substantial magnetic field presence at this distance has been reported in Refs.\cite{1989Natur.341..720K, 2015JCAP...01..011K}.
For each cluster, we considered $10^3$ realizations of the field, randomly varying the orientation of the magnetic field in each domain and randomly distributing the line-of-sight distance from the center of the cluster within 0 to 500 kpc. 

For each realization of the magnetic field, we calculate $P_{\gamma a}(E)$ for a set of ALP parameters ($m_a$, $\ga$) (blue lines in Figure~\ref{fig:conversion}).
We average this function over the described realizations and obtain the function $\bigl\langle P_{\gamma a}(E)\bigr\rangle$ -- black lines in Figure~\ref{fig:conversion}.
This function is well approximated by the shape, Eq.~\eqref{eq:p_ga} (red line in aforementioned Figure):
\begin{equation}
\bigl\langle P_{\gamma a} (E)\bigr\rangle = \frac{p_0}{1 + (E_c/E)^k}.
\end{equation}
This establishes a relation between the parameters ($p_0, E_c, k$) of the photon survival probability function $P_{\gamma\gamma}\equiv 1 - \bigl\langle P_{\gamma a}\bigr\rangle$ and the ALP parameters $(m_a, \ga)$, shown in Figure~\ref{fig:shape_parameters}.

\subsection{Dispersion of magnetic field strength.}
We repeated our analysis, spreading the values of $\beta$ and $\eta$ by up to $\pm$90\% around their adopted values. The characteristic energy $E_{c}$ and the plateau height $p_0$ remained constant within 5\%.
Variations in the distribution of domain sizes had a negligible impact on the averaged curves.

\begin{figure}[t!]
    \centering
    \includegraphics[width=0.7\textwidth]{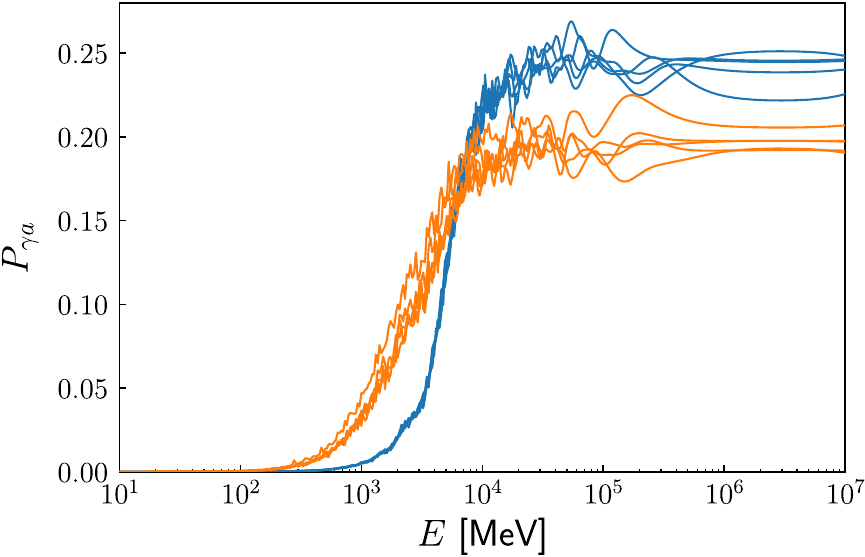}
    \caption{\textbf{Dependence of photon-to-ALP conversion probability on magnetic field parameters in clusters.} 
    Each curve shows the average probability of photon-to-ALP conversion over 29 realizations for the Coma cluster (\textbf{blue lines}) and Coma-like clusters (\textbf{orange lines}). For the blue curves, each realization varies the magnetic field direction within each domain and the distribution of domain sizes while maintaining a constant radial profile for the magnetic field, as specified by Eq.~\eqref{eq:magn_f}. The concentration of thermal electrons and magnetic field values in these realizations match those of the Coma cluster.
For the orange curves, in addition, the central amplitude of the magnetic field, $B_0$, is also varied. Specifically, $B_0$ is drawn from a log-uniform distribution ranging from $0.52$ to $52\,\mu {\rm G}$, with the mean level matching that of the Coma cluster. This variation in $B_0$ results in approximately a $20\%$ change in the average absorption feature. The ALP parameters are set to $m_a=\SI{3}{neV}$ and $\ga=\SI{2e-12}{\per\GeV}$, with domain sizes randomly varying between \SIrange{2}{34}{kpc}.}
    \label{fig:probs-different-clusters}
\end{figure}

As the photon-to-ALP conversion probability is directly proportional to the magnetic field value, we varied the parameter $B_0$ by an order of magnitude in each direction (from $\SI{0.52}{\mu G}$ to $\SI{52}{\mu G}$).
Figure~\ref{fig:probs-different-clusters} shows that in this case, $p_0$ varies by only 20\%, while the function $\bigl\langle P_{\gamma a}(E)\bigr\rangle$ maintains its shape~\eqref{eq:p_ga}.
This illustrates that averaging over the clusters with substantially different magnetic fields does not lead to the conversion probability dominated by the extreme values of the distribution but is primarily determined by the average magnetic field across the sample of clusters.

\subsection{Shifting the magnetic field amplitude.}
To estimate the potential impact of the change in the central value of $B_0$, we varied it from 3.1~$\mu$G to 6.5~$\mu$G.
The values of $B_0$ and $\eta$ are strongly correlated, with the smaller $\eta$ corresponding to smaller $B_0$. This correlation arises from the fact that the directly observed quantity is the rotation measure (RM):
\begin{equation}
\label{eq:rm}
RM = 812 \int\limits_{l.o.s.} n_e B_\parallel d\ell \quad\mbox{rad m}^{-2} \propto \frac{B_0}{3\beta(\eta+1)-1} 
\end{equation}
RM is sensitive to the mean value of the magnetic field along the line of sight (l.o.s.) and is measured with typically lower uncertainties than the derived parameters $B_0$ and $\eta$.
Therefore, we accompanied a change in $B_0$ by changing the slope $\eta$ between 0.4 and 0.7. 
Such a variation corresponds roughly to the 95\% CL ranges reported for the Coma cluster\cite{Bonafede:2010xg}. 
The associated uncertainty is shown as a green shaded region in Figure~\ref{fig:broad_constraints}.
It amounts to a change in $\ga$ for a fixed $m_a$ by about 20\%.

\begin{figure}[!t]
    \centering
    \includegraphics[width=0.7\textwidth]{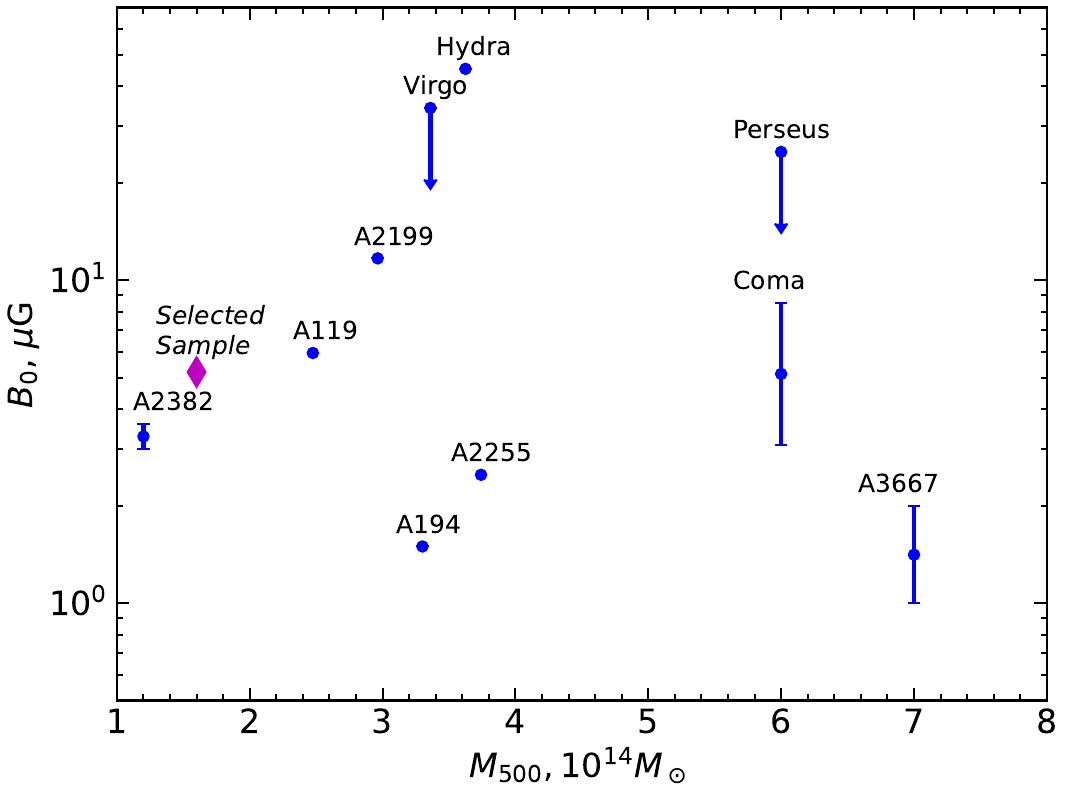}
    \caption{\textbf{Central magnetic field strength vs.\ mass for selected galaxy clusters.} 
    This plot shows $B_0$ in relation to the mass $M_{500}$ for a number of galaxy clusters, based on Refs.\cite{Bonafede:2010xg,1995A&A...302..680F,2006MNRAS.368.1500T,Cecil:2023iU,2019MNRAS.483..540L,2017A&A...603A.122G,2016A&A...594A..27P,2012MNRAS.427..550L,2008A&A...483..699G,2000MNRAS.313..617L,2001MNRAS.326....2T,2004NewAR..48.1145F,2022A&A...659A.146D}. The selected sample's log-average mass and adopted central magnetic field strength are highlighted in purple. The error bars indicate the uncertainties in the measurements of $B_0$, and the arrows indicate the upper limits of the magnetic field measurements.}
    \label{fig:scatter_plot}
\end{figure}

\subsection{Average magnetic field strength across the sample of clusters.}
A crucial factor in our analysis is the estimation of the average magnetic field strength in our sample of galaxy clusters. This estimation could potentially be biased due to a correlation between the central magnetic field strength ($B_0$) and the cluster mass ($M_{500}$). Our sample's log-average mass is $M_{500}\sim 1.6\times 10^{14}M_\odot$ (see Table~\ref{tab:agn_summary}), which is approximately four times smaller than the Coma cluster's mass of $M_{500}\sim 6\times 10^{14}M_\odot$ \cite{coma_mass}.

Current observational data do not reveal any obvious correlation between $B_0$ and $M_{500}$, as Figure~\ref{fig:scatter_plot} illustrates.
Moreover, different methods of evaluating the strength of the magnetic field (Faraday rotation, synchrotron diffuse radio emission, the inverse Compton hard X-ray emission) provide varying estimates of the magnetic field strength in clusters due to differences in measurement techniques, spatial scales, and the complex structure of magnetic fields in cluster environments, see, e.g.,\ the discussion in Refs.\cite{Feretti:2004vt,2012A&ARv..20...54F}.
Furthermore, an analysis comparing clusters with high and low temperatures reveals no significant variations in the RM data \cite{2010A&A...522A.105G}. These findings, therefore, indicate the absence of a strong relation between the magnetic field and the cluster mass, given the well-established mass–temperature relation for clusters of galaxies, see, e.g.,\ Ref\cite{Finoguenov:2000qb}.
Given this lack of a discernible trend, we argue that the magnetic field profile of the Coma cluster (with $B_0\sim 5.2\,\mu$G) can be considered representative of our entire cluster sample.

Recent N-body simulations\cite{2024A&A...686A.157N} suggest, however, a scaling relation of $B_0\sim M_{500}^{1/3}$. If this relation holds, the average $B_0$ in our sample would be approximately $1.6$ times lower than in massive clusters like Coma.
To assess the potential impact of this scaling on our results, we performed an additional analysis. We explicitly down-scaled the magnetic field in our $\langle P_{\gamma a}(E)\rangle$ calculations based on the $M_{500}$ masses of clusters in our sample (see Table~\ref{tab:agn_summary}) and the aforementioned scaling. 
The resulting limits on ALP parameters are shown as a blue dash-dot-dotted line in Figure~\ref{fig:broad_constraints}. These limits are a factor of $1.6$ weaker than those obtained using the characteristic magnetic field profile.

We note that the dependency of the central magnetic field on the redshift\cite{2022A&A...665A..71O,2024A&A...686A.157N} of the clusters can be neglected, as the clusters are located at low ($z_{\text{\sc gc}}\lesssim 0.4$) redshifts.

\subsection{Correction for the finite sample size.}
\begin{figure}[t!]
    \centering
    \includegraphics[width=0.7\textwidth]{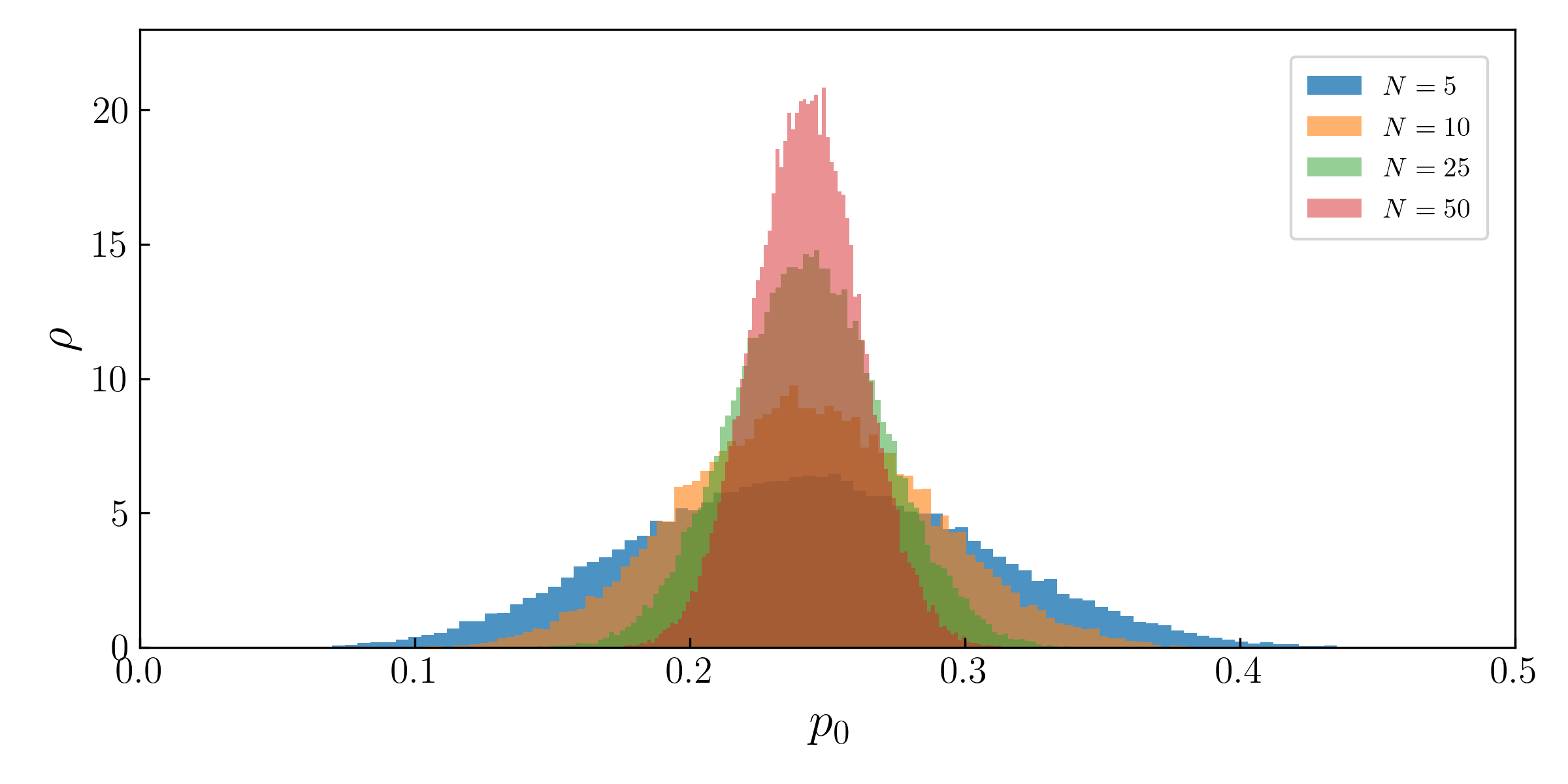}
    \caption{\textbf{Asymptotic value of $P_{\gamma a}$ at high energies depending on the number of realizations $N$}. Probability density, $\rho$, of finding a certain value of $p_0$. 
    All results are for  $(m_a, \ga) = (\SI{3}{\nano\eV}, \SI{2e-12}{\per\GeV})$.}
    \label{fig:step-distribution}
\end{figure}

Our sample size is much smaller than the simulated number of the realizations of the magnetic field. 
Figure~\ref{fig:step-distribution} illustrates the dependency of the parameter $p_0$ (Eq.~\eqref{eq:p_ga})
on the size of the sample. 
As expected, the scatter drastically reduces with the number of considered objects. 
However, the width of the distribution in Figure~\ref{fig:step-distribution} is non-negligible for the sample of 29 objects. 
Therefore, taking the central value of $p_0$ could lead to an overestimate of the strength of the exclusions on $\ga$.
To address this issue, we allowed $p_0$ to vary within the $\pm 1\sigma$ interval of the corresponding distribution (see Figure~\ref{fig:step-distribution}), i.e., with $\pm 20\%$ from the central value. 
For 29 objects, the log-width of the distribution remains roughly constant with $(m_a, g_a)$, which allowed us to use the same range of variations of $p_0$ for all ALP parameters.

\subsection{Selection of AGN-cluster pairs.}
\label{sec:selection}
We identified $\gamma$-ray bright AGNs located beyond or within known galaxy clusters based on the most recent catalog of the high-altitude ($|b|>10^\circ$) sources, 4LAC-DR3-h\cite{4lacdr3}, and the catalog of the optically identified galaxy clusters\cite{2018MNRAS.475..343W}. 
Among 1,806 AGNs with known redshifts and emission in the GeV range, and 47,600
clusters with known redshifts, we were able to identify 29 AGN-cluster pairs for which the line of sight to the AGN passes through the cluster at a comoving distance not exceeding 500~kpc\cite{2018MNRAS.475..343W, 2006A&A...446..429P} and $z_{\text{\sc agn}} \ge z_{\text{\sc gc}}$. We additionally included in the sample two nearby AGNs (NGC~1275 and M87) located within the Perseus and Virgo clusters, respectively. 
The basic properties of the AGNs/galaxy clusters sample are summarized in Table~\ref{tab:agn_summary}.

\subsection{Data and data analysis.}
\label{sec:data}
The AGN spectra are provided by the \lat collaboration as part of the 4FGL-DR4 catalog\cite{4fgldr4, 2022ApJS..260...53A} and correspond to 14-year time-averaged spectra. For each object from the selected sample, we considered its \lat spectral energy distribution (SED) in 8 energy bins as reported in the 4FGL catalog. We assumed also that, in addition to the statistical uncertainty, the spectral points are characterized by a certain level of systematic uncertainty (added in quadrature). We consider two choices of systematic uncertainty\cite{fermi_lat_website}: \textit{(i)} optimistic (systematic set to 0); \textit{(ii)} ``nominal'' (3\% systematics at all energies except $E<100$~MeV and $E>100$~GeV, for which it is 10\%). 
In this work, we present the results for each of these choices.

We fit the AGN spectra with the ``baseline'' EBL-corrected log-parabola models, 
\begin{align}
& \frac{dN}{dE} = N_0 \left(E/{E_0}\right)^{-\alpha-\beta \log(E/E_0) } \cdot\kappa_{\text{\sc ebl}}(E, z_{\text{\sc agn}})\;,
\end{align}
where normalization $N_0$ and spectral parameters $\alpha$ and $\beta$ are the free fitting parameters, and $E_0 =\SI{1}{GeV}$. The EBL-correction factor $\kappa_{\text{\sc ebl}}(E,z)$ was calculated for AGN redshift $z_{\text{\sc agn}}$ with the help of the absorption model provided within the \texttt{naima} Python module\cite{naima} based on adopted EBL model\cite{dominguez11}.

\begin{figure}[t!]
    \centering
    \includegraphics[width=0.7\textwidth]{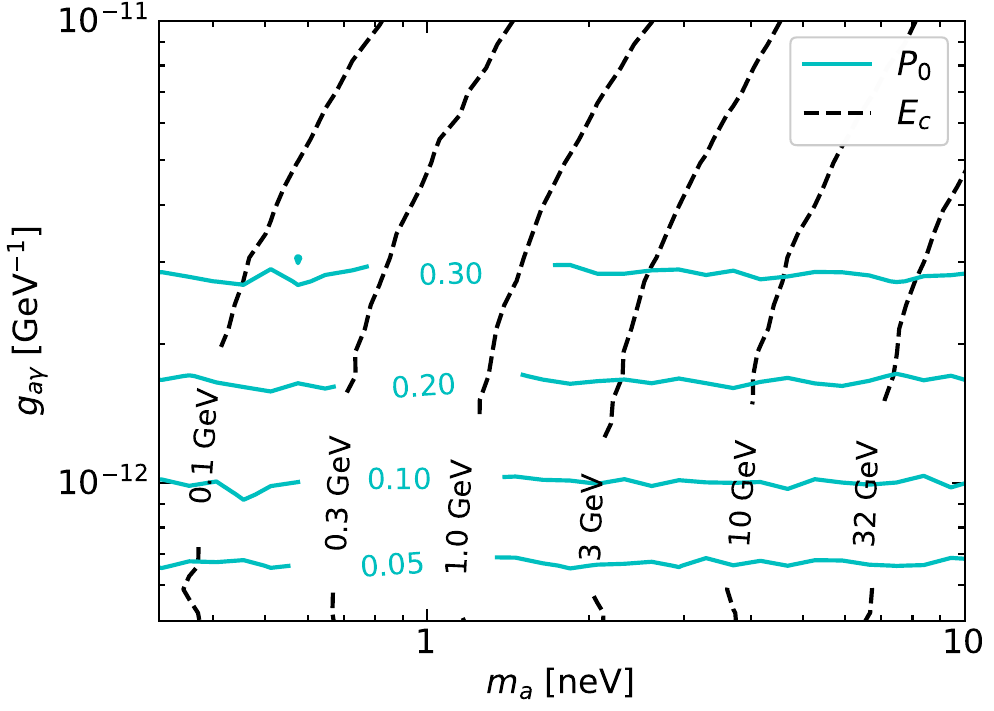}
    \caption{\textbf{Dependence of the averaged photon survival probability function's shape parameters on ALP parameters.} Dependence of the shape parameters $E_c$ and $p_0$, as defined by Equation~\eqref{eq:p_ga}, on the ALP parameters $(m_a,\ga)$ for the range of parameters considered in this work.}
    \label{fig:shape_parameters}
\end{figure}

Aiming to probe photon-to-ALP conversion as the AGN photons propagate through the clusters of galaxies,
for a range of ALP masses and coupling constants, we have considered an ``ALP-model''
\begin{align}
& \frac{dN}{dE} = N_0 \left(E/E_0\right)^{-\alpha-\beta \log(E/E_0) } \cdot\kappa_{\text{\sc ebl}}(E, z_{\text{\sc agn}})\cdot P_{\gamma\gamma}
\end{align}
where $P_{\gamma\gamma}$ is given by Eq.~(\ref{eq:p_ga}). 
Three extra parameters of the function $P_{\gamma\gamma}$ ($p_0$, $E_c$,  $k$) are related to the ALP parameters ($m_a$ and $\ga$), as discussed above. The dependency of $p_0$ and $E_c$ on ALP parameters is shown in Figure~\ref{fig:shape_parameters}.

\begin{figure}[t]
    \centering
    \includegraphics[width=0.7\textwidth]{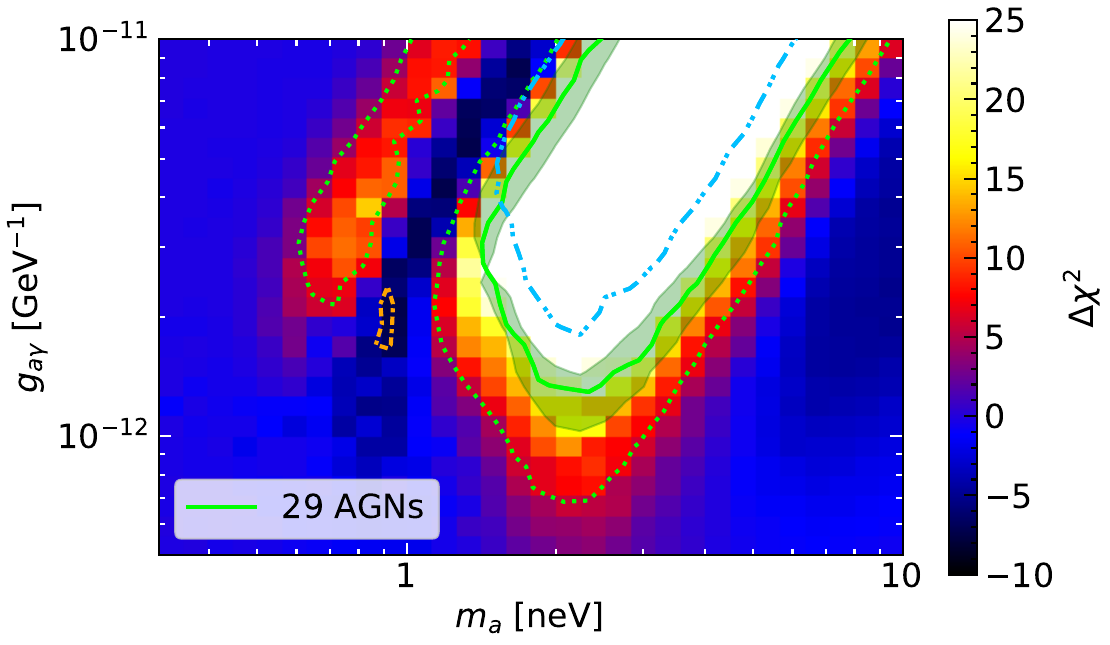}
    \caption{\textbf{$\chi^2$ change and ALP exclusion regions}. 
    The colors illustrate the change in $\chi^2$ between the baseline and the ALP models for the combined fit of 29 AGNs, assuming statistical-only uncertainties. The green contours represent $2\sigma$-excluded regions ($\Delta\chi^2=6.2$ for 2 d.o.f.) for two different treatments of the systematic uncertainties. Dotted and solid lines correspond to the zero (i.e., statistical only) and nominal levels of \lat data systematic uncertainties.
    The shaded region shows the uncertainty related to the magnetic field estimation in the Coma cluster. The dash-dot-dotted region illustrates the weakening of the limits for the maximally conservative choice of the magnetic field in the selected sample of galaxies (for the nominal level of \lat systematics). The dash-dotted orange region corresponds to a $2\sigma$ detection region where the ALP model actually improves $\Delta\chi^2$ by at least $6.2$.}
    \label{fig:chi2_agns}
\end{figure}

The difference between the joint best-fit $\chi^2$ for the baseline and ALP-model (allowing $p_0$ variations within $\pm 20$\% around the mean value to correct for the finite size of the clusters' sample) is shown in Figure~\ref{fig:chi2_agns}. 
The green contours in this figure represent a deterioration in the fit with the ALP-model by $\Delta \chi^2 = 6.2$ with respect to the baseline model, corresponding to a $2\sigma$ excluded region for 2 degrees of freedom (d.o.f.). The green dotted line indicates the limits for the statistical-only uncertainty (0\% systematics). The green solid line corresponds to the nominal \lat systematics. The shaded region corresponds to the variations of limits derived for the nominal level of \lat systematics for the different profiles of the Coma cluster's magnetic field (see above). The dash-dot-dotted region indicates the weakening of the limits in case the magnetic field in each of the clusters in the sample is scaled with respect to the mass of the cluster, as discussed above. These same contours are depicted in Figure~\ref{fig:broad_constraints}.

The orange dot-dashed region corresponds to the region where ALPs are detected with a $\gtrsim 2\sigma$ significance ($\Delta\chi^2\leq -6.2$) in the absence of systematic uncertainty (purely statistical bound). 
The maximal improvement of the fit corresponds to $\Delta\chi^2 \approx -7.4$ for $m_a \approx \SI{1}{\nano\eV}$ and $\ga \approx 2 \times 10^{-12}$GeV$^{-1}$. See Table~\ref{tab:detection_table} for a summary of the $\Delta\chi^2$ improvement in individual objects. 
We note that this detection is not statistically significant and disappears in the presence of systematic uncertainties.

\subsection{CTAO sensitivity.}
To estimate the sensitivity of the forthcoming TeV CTAO for similar studies we simulated a similar sample of 29 AGNs. We assumed that the AGN spectra continue as power laws in the energy band $0.03-10$~TeV and that the CTAO will be able to measure 8 spectral points in this energy band. We further assume that the uncertainties of the flux measurements are dominated by 10\% systematic uncertainties. We repeat the procedure described above for ALP searches in the simulated CTAO-only dataset. The estimated level of exclusions derived from such a dataset is shown with a red dot-dashed line in Figure~\ref{fig:broad_constraints}.

\subsection{Consistency checks.}
\label{sec:checks}
To verify the self-consistency and statistical reliability of our proposed approach, we artificially added to the data the ALP-induced suppression and then tried to detect it, using our procedure. 
This has been done for several values of $(m_a, \ga)$ within the excluded region (the solid green contour in Fig.~\ref{fig:chi2_agns}).
Namely, using the \texttt{ALPro} code, we generated $10^3$ random photon-to-ALP conversion probability curves for several ALP parameters.

We then randomly drew 29 conversion curves, binned them to match \lat spectral binning, and applied them to the AGN spectra. 
The procedure was repeated, creating $10^3$ samples of 29 AGNs, each ``containing'' the ALP with the same parameters $(m_a, \ga)$.

\begin{figure}[t]
    \centering
    \includegraphics[width=0.42\textwidth]{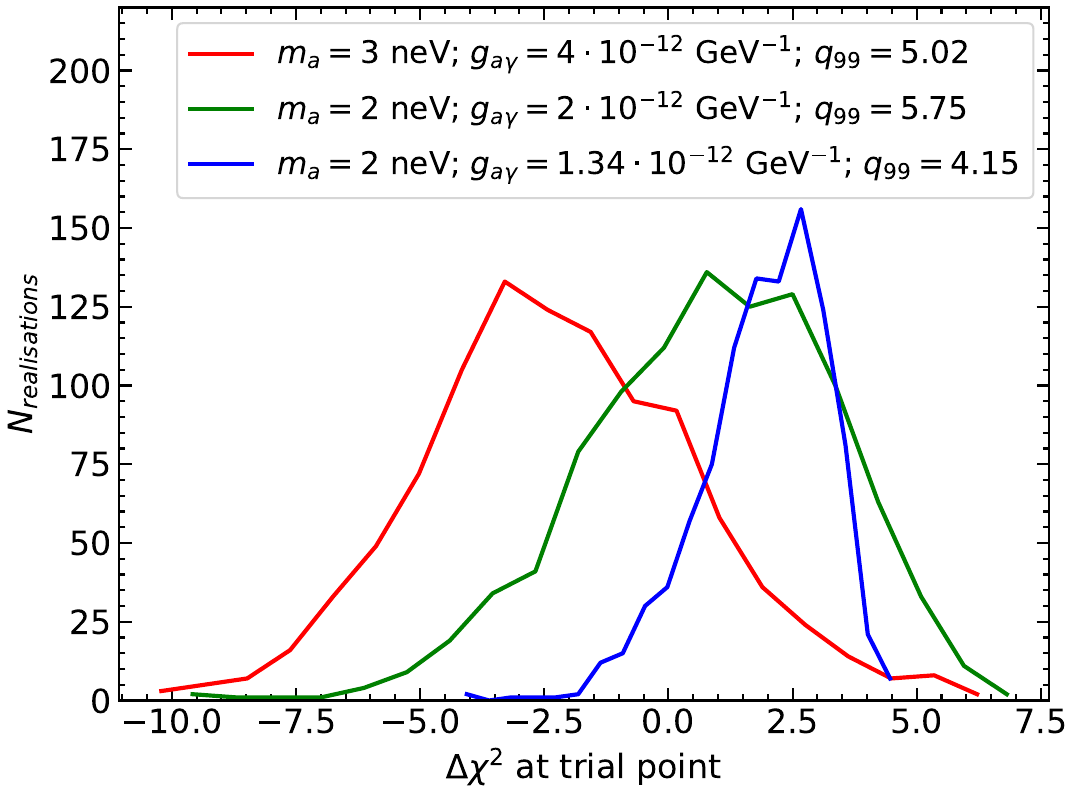}
    \includegraphics[width=0.55\textwidth]{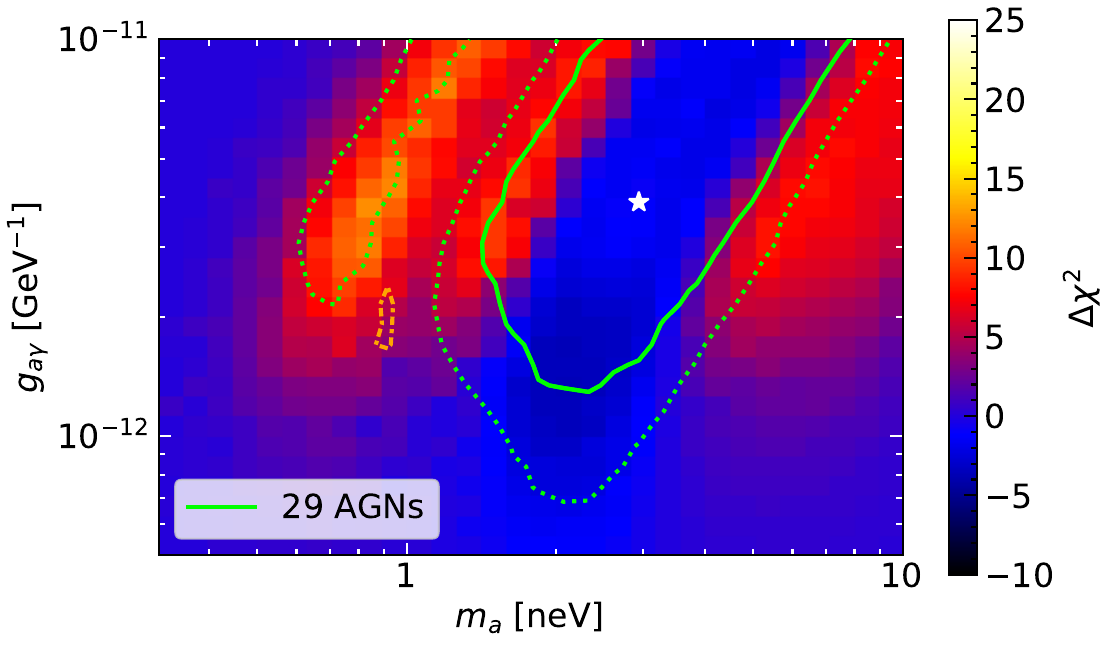}
    \caption{\textbf{Consistency checks: adding ALP to the data.} Left panel: Distribution of $\chi^2$ values for several ALP masses and coupling constants computed using random ALP absorption curves convolved with the data. $\Delta\chi^2$ for the parameters corresponding to the red, green, and blue curves for the unmodified data were 23.87, 12.74, and 6.4, respectively. 
    Right panel: Simulated exclusion plane for an ALP with specific parameters $m_a=\SI{3}{neV}$, $\ga=\SI{4e-12}{\per\GeV}$ depicted in the plot with a white star. The green contours show the exclusion regions identical to those in Figure~\ref{fig:chi2_agns}. For both panels, the nominal level of \lat systematics was considered.}
    \label{fig:trials}
\end{figure}

We applied our detection procedure (assuming the nominal level of \lat systematics) to each sample to determine the frequency of falsely excluding the ALP. Figure~\ref{fig:trials} (left panel) shows the resulting distributions of $\Delta \chi^2$ values. 
We found that for all selected trial points, more than 99\% of the $\Delta\chi^2$ values are below 6.2 (see $q_{99}$ values in the legend).
This demonstrates the robustness of our approach, showing that if an ALP with specific parameters (as shown in the inset) is present in the data, our method does not falsely exclude these parameters.

The right panel of Figure~\ref{fig:trials} illustrates the ``exclusion plot'' based on our sample modified by the photon-to-ALP conversion for an ALP with parameters $m_a=\SI{3}{\nano\eV}$, $\ga=\SI{4e-12}{\per\GeV}$ (i.e., a randomly drawn realization from the red curve in the left panel). 
These ALP parameters are indicated with a white star. The green contours are identical to the exclusion contours in Figure~\ref{fig:chi2_agns}. Notably, the ALP with the corresponding parameters is no longer excluded and, moreover, it creates an ``island'' in the parameter space similar to that seen in the real data when purely statistical uncertainties are considered. 

\begin{table}[]
    \centering
    \begin{tabular}{c|c|c|c|c|c|c}
    \hline
 \lat source  &  $z_{\text{\sc agn}}$ & Cluster  & $z_{\text{\sc gc}}$  & D  & $M_{500}$ &Reference \\
   &   &   &   & (kpc)  & ($10^{14}M_\odot$) &\\
   \hline
4FGL J0014.2+0854 &  0.163 & J001419.7+085401 & 0.159 & 192.8 & 2.9 & T1 \cite{2018MNRAS.475..343W}\\
4FGL J0038.2-2459 &  1.196 & J003757.6-250425 & 0.064 & 478.8 & 1.9 & T1 \cite{2018MNRAS.475..343W}\\
4FGL J0049.0+2252 &  0.264 & J004857.8+225427 & 0.151 & 392.3 & 1.1 & T1 \cite{2018MNRAS.475..343W}\\
4FGL J0132.7-0804 &  0.148 & J013241.2-080405 & 0.136 & 130.0 & 2.4 & T1,T3 \cite{2018MNRAS.475..343W}\\
4FGL J0317.8-4414 &  0.076 & J031757.7-441418 & 0.061 & 58.4 & 3.0 & T1,T3 \cite{2018MNRAS.475..343W}\\
4FGL J0617.7-1715 &  0.098 & J061733.4-171525 & 0.093 & 237.1 & 2.8 & T1 \cite{2018MNRAS.475..343W}\\
4FGL J0912.5+1556 &  0.212 & J091230.5+155658 & 0.196 & 254.5 & 1.1 & T1 \cite{2018MNRAS.475..343W}\\
4FGL J0914.4+0249 &  0.427 & J091428.2+025036 & 0.160 & 141.2 & 2.0 & T1 \cite{2018MNRAS.475..343W}\\
4FGL J1010.8-0158 &  0.896 & J101056.5-015927 & 0.196 & 463.8 & 4.3 & T1 \cite{2018MNRAS.475..343W}\\
4FGL J1013.7+3444 &  0.208 & J101350.8+344251 & 0.143 & 277.9 & 2.3 & T1 \cite{2018MNRAS.475..343W}\\
4FGL J1058.4+0133 &  0.890 & J105811.0+013617 & 0.038 & 254.1 & 2.0 & T1,T3 \cite{2018MNRAS.475..343W}\\
4FGL J1202.5+3852 &  0.805 & J120230.5+385219 & 0.283 & 350.9 & 1.1 & T1 \cite{2018MNRAS.475..343W}\\
4FGL J1213.0+5129 &  0.796 & J121246.7+513250 & 0.084 & 467.4 & 0.9 & T1 \cite{2018MNRAS.475..343W}\\
4FGL J1303.0+2434 &  0.993 & J130303.4+243456 & 0.295 & 135.0 & 2.6 & T1 \cite{2018MNRAS.475..343W}\\
4FGL J1353.2+3740 &  0.216 & J135314.1+374114, & 0.216 & 95.1 & 1.8 & T1 \cite{2018MNRAS.475..343W}\\
                  &        & J135318.5+373838 & 0.094 & 291.7 & 1.9 & T1 \cite{2018MNRAS.475..343W}\\
4FGL J1508.8+2708 &  0.270 & J150843.1+271046 & 0.081 & 269.7 & 1.4 & T1 \cite{2018MNRAS.475..343W}\\
4FGL J1516.8+2918 &  0.130 & J151641.6+291809 & 0.130 & 339.6 & 1.2 & T1,T3 \cite{2018MNRAS.475..343W}\\
4FGL J1615.6+4712 &  0.199 & J161541.3+471004 & 0.198 & 498.5 & 2.3 & T1 \cite{2018MNRAS.475..343W}\\
4FGL J2041.9-3735 &  0.099 & J204154.9-373849 & 0.093 & 413.9 & 1.2 & T1 \cite{2018MNRAS.475..343W}\\
4FGL J2314.0+1445 &  0.164 & J231357.4+144423 & 0.143 & 237.9 & 2.2 & T1 \cite{2018MNRAS.475..343W}\\
4FGL J2321.9+2734 &  1.253 & J232159.1+273443 & 0.093 & 53.0 & 1.9 & T1,T3 \cite{2018MNRAS.475..343W}\\
4FGL J2336.6+2356 &  0.127 & J233642.1+235529 & 0.105 & 205.0 & 2.3 & T1,T3 \cite{2018MNRAS.475..343W}\\
4FGL J2338.9+2124 &  0.291 & J233853.3+212753 & 0.071 & 326.3 & 1.3 & T1 \cite{2018MNRAS.475..343W}\\
\hline
4FGL J0303.3-7913 &  1.115 & J030351.6-791228 & 0.189 & 383.8 & 0.3 & T2 \cite{2018MNRAS.475..343W}\\
4FGL J0309.4-4000 &  0.193 & J030937.3-400045 & 0.141 & 353.4 & 0.2 & T2 \cite{2018MNRAS.475..343W}\\
4FGL J0654.4+4514 &  0.928 & J065427.7+451447 & 0.374 & 174.7 & 0.6 & T2 \cite{2018MNRAS.475..343W}\\
4FGL J1242.9+7315 &  0.075 & J124311.3+731559 & 0.074 & 87.7 & 0.3 & T2 \cite{2018MNRAS.475..343W}\\
\hline
4FGL J0319.8+4130 &  0.017559 & Perseus & 0.017559 & -- & 6.1 & NED\cite{perseus_mass}\\
4FGL J1230.8+1223 & 0.004283 & Virgo & 0.004283 & -- &3.36 & NED\cite{virgo_mass} \\
\hline
    \end{tabular}
    \caption{List of \lat AGNs located behind/within galaxy clusters. The table summarizes the names of AGNs as presented in the 4FGL \lat catalog, their redshifts $z_{\text{\sc agn}}$, corresponding clusters, and their redshifts $z_{\text{\sc gc}}$. The comoving distance at the redshift of the cluster, corresponding to the angular separation between the AGN and the cluster's center, is given by $D$. The $M_{500}$ of the clusters\cite{2018MNRAS.475..343W} is estimated as $M_{500}\sim 0.5\times 10^{14}M_\odot (R_{L*}/8.0)^{1.08}$.   
    The reference column specifies the source from which the clusters' data were adopted.}
    \label{tab:agn_summary}
\end{table}

\begin{table}[]
    \centering
    \begin{tabular}{c|c|c|c|c}
    \hline
Fermi/LAT source & Average significance & $\chi^2_{0}$ & $\chi^2_{\text{ALP}}$ & $\Delta\chi^2$ \\
\hline
4FGL J0014.2+0854 & 5.97  & 2.04 & 2.22 & 0.18  \\
4FGL J0038.2-2459 & 97.56  & 16.03 & 13.75 & -2.28  \\
4FGL J0049.0+2252 & 8.03  & 0.99 & 0.97 & -0.02  \\
4FGL J0132.7-0804 & 9.97  & 2.16 & 2.24 & 0.08  \\
4FGL J0303.3-7913 & 10.82  & 3.75 & 3.49 & -0.26  \\
4FGL J0309.4-4000 & 7.53  & 0.89 & 1.05 & 0.15  \\
4FGL J0317.8-4414 & 5.78  & 1.64 & 1.56 & -0.09  \\
4FGL J0617.7-1715 & 29.06  & 4.89 & 3.85 & -1.04  \\
4FGL J0654.4+4514 & 36.34  & 3.64 & 3.64 & 0.01  \\
4FGL J0912.5+1556 & 12.90  & 3.02 & 3.08 & 0.06  \\
4FGL J0914.4+0249 & 6.56  & 1.35 & 1.20 & -0.14  \\
4FGL J1010.8-0158 & 9.64  & 1.90 & 1.98 & 0.08  \\
4FGL J1013.7+3444 & 20.74  & 5.06 & 5.42 & 0.36  \\
4FGL J1058.4+0133 & 102.24  & 2.88 & 3.17 & 0.29  \\
4FGL J1202.5+3852 & 10.22  & 1.82 & 1.86 & 0.03  \\
4FGL J1213.0+5129 & 16.84  & 4.72 & 4.49 & -0.23  \\
4FGL J1242.9+7315 & 12.31  & 7.80 & 7.96 & 0.16  \\
4FGL J1303.0+2434 & 53.97  & 2.07 & 1.53 & -0.54  \\
4FGL J1353.2+3740 & 9.85  & 2.71 & 2.56 & -0.14  \\
4FGL J1508.8+2708 & 12.74  & 0.81 & 0.76 & -0.05  \\
4FGL J1516.8+2918 & 7.66  & 1.36 & 1.31 & -0.05  \\
4FGL J1615.6+4712 & 20.86  & 2.86 & 2.42 & -0.44  \\
4FGL J2041.9-3735 & 14.21  & 0.94 & 0.96 & 0.02  \\
4FGL J2314.0+1445 & 22.79  & 4.78 & 4.72 & -0.06  \\
4FGL J2321.9+2734 & 36.10  & 4.85 & 4.92 & 0.07  \\
4FGL J2336.6+2356 & 12.05  & 5.92 & 5.80 & -0.13  \\
4FGL J2338.9+2124 & 17.84  & 1.17 & 1.53 & 0.36  \\
4FGL J1230.8+1223 & 46.98  & 9.73 & 10.78 & 1.06  \\
4FGL J0319.8+4130 & 251.54  & 7.59 & 2.80 & -4.79  \\
\hline
Total & -- & 109.38 & 102.02 & -7.36\\
\hline
    \end{tabular}
    \caption{A summary of the best-fit $\chi^2$ of the models fitting the spectra of selected \lat sources with ($\chi^2_{\text{ALP}}$) and without ($\chi^2_{0}$) ALP components; see the text for more details. The mass and coupling constant of the ALP were selected to match the parameters of the marginal $2\sigma$ detection ($m_a=9.1\times 10^{-10}$~eV, $\ga=2.1\times 10^{-12}$~GeV$^{-1}$). The negative $\Delta\chi^2=\chi^2_{\text{ALP}}-\chi^2_{0}$ indicates an improvement of the ALP-invoking model compared to the baseline fit model.
    The ``Average significance'' column indicates the average significance of the given source according to the 4FGL catalog.}
    \label{tab:detection_table}
\end{table}
\begin{addendum}

\item[Acknowledgments.]
We would like to thank our colleagues D.~R.~Harvey, D.~F.~G.~Fiorillo, J.~U.~Fynbo,
M.~Kachelrie{\ss}, and M.~Meyer  for reading the draft of this manuscript and providing important comments that helped to improve the quality of this work. We greatly appreciate their insightful feedback and constructive suggestions.

The authors acknowledge support by the state of Baden-W\"urttemberg through bwHPC.  
LZ's work is supported by a Scholars at Risk Denmark Fellowship for Scholars from Ukrainian Universities (SARU Fellowship) at the University of Copenhagen. 
This project has received funding through the MSCA4Ukraine project, grant number Ref 1.4 - UKR - 1245772 - MSCA4Ukraine, which is funded by the European Union. Views and opinions expressed are however those of the author(s) only and do not necessarily reflect those of the European Union, the European Research Executive Agency or the MSCA4Ukraine Consortium. Neither the European Union nor the European Research Executive Agency, nor the MSCA4Ukraine Consortium as a whole nor any individual member institutions of the MSCA4Ukraine Consortium can be held responsible for them.
This research was conducted with support from the Centre for the Collective Use of Scientific Equipment "Laboratory of High Energy Physics and Astrophysics" of Taras Shevchenko National University of Kyiv. 

\item[Data Availability.] The data from \lat observations used in this paper is taken from the publicly available 4FGL-DR4 catalog\cite{4fgldr4}. The active galactic nuclei sample is adopted from the catalog of high-altitude ($|b|>10^\circ$) objects 4LAC-DR3-h\cite{4lacdr3} and the catalog of clusters identified from all-sky surveys of 2MASS, WISE, and SuperCOSMOS\cite{2018MNRAS.475..343W}. 

\item[Code Availability.] The ALP propagation code used in this study is a modified version of the publicly available code \texttt{AlPro} (Axion-Like PROpagation)\cite{Matthews2022}, available at \url{https://github.com/jhmatthews/alpro}.

\item[Competing Interests Statement] The authors declare no competing interests.

\item[Authors contribution.] All authors have analyzed and interpreted the data and prepared the manuscript. All authors have reviewed, discussed, and commented on the present results and on the manuscript. 
\end{addendum}
\bigskip


\end{document}